\documentclass[a4paper,12pt]{extarticle}
\usepackage{graphicx}
\usepackage{preamble}
\usepackage{setspace}
\usepackage[font=small,labelfont=bf]{caption}
\usetikzlibrary{calc}
\usepackage{amsmath,amssymb}
\usetikzlibrary{decorations.pathmorphing}

\definecolor{refkey}{rgb}{0.9451,0.2706,0.4941}
\definecolor{labelkey}{rgb}{0.9451,0.2706,0.4941}
\usepackage{cite}

\def\z2{$\mathbb{Z}_2$}

\definecolor{darkgray}{rgb}{0.33, 0.33, 0.33}
\usepackage{braket}

\newcommand{\Tr}{{\mathrm{Tr}}}

\usepackage[left=30mm,right=30mm,top=30mm,bottom=30mm]{geometry}
\usetikzlibrary{arrows.meta,calc}
\usepackage{authblk}
\numberwithin{equation}{section}
\definecolor{MyGreen}{RGB}{10,120,10}
\usetikzlibrary{shapes.misc}
\tikzset{cross/.style={cross out, draw=black, minimum size=2*(#1-\pgflinewidth), inner sep=0pt, outer sep=0pt},
cross/.default={1pt}}
\usetikzlibrary{positioning}
\usepackage{mdframed}

\def\bZ{\mathbb{Z}}
\def\cH{\mathcal{H}}

\usetikzlibrary{shapes.arrows}

\begin{document}
\onehalfspacing
\title{
What happens to wavepackets of fermions
when scattered by the Maldacena-Ludwig wall?
}

\author[1]{Yuji Tachikawa\footnote{\href{mailto:yuji.tachikawa@ipmu.jp}{yuji.tachikawa@ipmu.jp}}}
\author[1]{Keita Tsuji\footnote{\href{mailto:keita.tsuji@ipmu.jp}{keita.tsuji@ipmu.jp}}}
\author[2]{Masataka Watanabe\footnote{\href{mailto:max.washton@gmail.com}{max.washton@gmail.com}}}
\affil[1]{\it \small Kavli Institute for the Physics and Mathematics of the Universe (WPI),\par
The University of Tokyo, Chiba 277-8583, Japan}
\affil[2]{\it \small 
Faculty of Science,
The University of Tokyo, Tokyo 113-0033, Japan}

\date{}
\pagenumbering{gobble}

\maketitle

\thispagestyle{empty}
\begin{abstract}
\noindent
We study wavepackets of exotic excitations after two-dimensional fermions are scattered by the boundary condition constructed by Maldacena and Ludwig,
which turns elementary excitations into  exotic fractionally-charged objects.
They are of interest in the s-wave approximation of 
the fermion-monopole scattering in four-dimensional QED and of the multi-channel Kondo effect.
We in particular give an explicit expression of the outgoing state of a pair of such particles;
we then examine its properties, such as the charge density $\braket{J(x)}$ and the expectation value
$\braket{N}$ of the number of fermions and anti-fermions in the state.
The charge density $\braket{J(x)}$  is found to be  localized with its integral  finite and fractional,
while the expectation value $\braket{N}$ diverges when the wavepacket is localized to a point.

\end{abstract}
\newpage
\pagenumbering{arabic}
\pagestyle{plain}
\tableofcontents



\setcounter{tocdepth}{2}

\section{Introduction}

\label{sec:intro}

When Callan and Rubakov first studied the scattering of charged fermions by a monopole in \cite{Callan:1982ac,Rubakov:1982fp},
they found that the outgoing wave has charges which are in general incompatible 
with those of the standard elementary excitations of the system.
For brevity, let us call such excitations \emph{exotic}.

The issue is already visible in the s-wave approximation.
Suppose that there are $N_f$ two-dimensional complex fermions 
in the effective two-dimensional spacetime spanned by the time direction $t$
and the radial direction $r$.
The symmetry of the system dictates that the boundary condition at $r=0$
is such that the diagonal $U(1)$ charge is flipped,
while the $SU(N_f)$ charge is kept fixed. 
This means that the outgoing wave is necessarily exotic
if we inject a single elementary fermion toward the boundary $r=0$
when $N_f>2$.

The same issue arises in the analysis of the multi-channel Kondo effect 
in condensed matter theory, where the scattering of electrons by a magnetic impurity is studied.
The s-wave approximation similarly results in a boundary condition at $r=0$
which reflects an incoming electron into something with exotic charges.

The low-energy boundary conditions in both cases of the Callan-Rubakov effect
and of the multi-channel Kondo effect are known to be essentially the same,
and were determined already three decades ago by the method of conformal field theory \cite{Callan:1994ub,Maldacena:1995pq,Affleck:1995ge}.
This allowed the computations of scattering amplitudes 
and many other physical properties of the system,
but the nature of the exotic outgoing excitations continues 
to perplex and fascinate us,
resulting in a flurry of works studying this issue in the last several years.\footnote{%
For example, see e.g.~\cite{Tong:2019bbk,Smith:2020rru,Smith:2020nuf,Fukusumi:2021zme,Loladze:2024ayk,Loladze:2025jsq} for the studies in two dimensions
and ~\cite{Csaki:2020inw,Kitano:2021pwt,Csaki:2021ozp,Brennan:2021ewu,Hamada:2022eiv,vanBeest:2023dbu,Brennan:2023tae,vanBeest:2023mbs,Khoze:2024hlb,Csaki:2024ajo,Bolognesi:2024kkb,Bogojevic:2024xtx} for the studies in four dimensions, 
both in the context of high energy physics theory,
and \cite{Ljepoja:2023nnz,Ljepoja:2023krz} in the context of condensed matter theory.
There are also proposals \cite{Landau:2017kpw,Sela:2023evy} 
to observe experimental signatures of exotic excitations in the context of Kondo physics,
and a numerical simulation of analogous exotic scattering processes \cite{Ueda:2025ecm}.}
In this paper we would like to shed light on this question from a slightly different angle,
by studying the \emph{state} of the wavepackets after the scattering,
mainly in the two-dimensional setup.

\begin{figure}
\centering
\begin{tikzpicture}[line cap=round,line join=round]
  \pgfmathsetmacro{\X}{1.5}
  \pgfmathsetmacro{\Y}{1.5}
  \pgfmathsetmacro{\h}{0.2}

  \draw[black!30, line width = \h * 1cm] (0, -\Y) -- (0, \Y);
  \draw[->, >=stealth] (0, -\Y) -- ++(0, \Y * 2) node[above left] {$t$};
  \draw[->, >=stealth] (0, 0) -- ++(\X, 0) node[below right] {$r$};
  
  \draw[->, >=stealth, red, semithick] (0.6, -0.6) -- (0.3, -0.3);
  \draw[red, semithick] (0.6, -0.8)
    .. controls ++(0.2, 0) and ++(-0.1, 0) ..
    ++(0.2, 0.2)
    .. controls ++(0.1, 0) and ++(-0.2, 0) ..
    ++(0.2, -0.2);
  \draw[->, >=stealth, blue, semithick] (0.3, 0.3) -- (0.6, 0.6);
  \draw[blue, semithick] (0.6, 0.8)
    .. controls ++(0.2, 0) and ++(-0.1, 0) ..
    ++(0.2, 0.2)
    .. controls ++(0.1, 0) and ++(-0.2, 0) ..
    ++(0.2, -0.2);

  \node [draw=black, single arrow, single arrow head extend=2mm, minimum height=8mm] at (3.5, 0) {};
  \node at (3.5, -0.6) {unfold};

  \begin{scope}[shift={(6.5,0)}]
    \draw[black!30, line width = \h * 1cm] (0, -\Y) -- (0, \Y);
    \draw[->, >=stealth] (0, -\Y) -- ++(0, \Y * 2) node[above left] {$t$};
    \draw[->, >=stealth] (-\X, 0) -- ++(\X*2, 0) node[below right] {$x$};

    \draw[->, >=stealth, red, semithick] (-0.6, -0.6) -- (-0.3, -0.3);
    \draw[red, semithick] (-0.9, -0.9)
      .. controls ++(0.2, 0) and ++(-0.1, 0) ..
      ++(0.2, 0.2)
      .. controls ++(0.1, 0) and ++(-0.2, 0) ..
      ++(0.2, -0.2);
    \draw[->, >=stealth, blue, semithick] (0.3, 0.3) -- (0.6, 0.6);
    \draw[blue, semithick] (0.6, 0.8)
      .. controls ++(0.2, 0) and ++(-0.1, 0) ..
      ++(0.2, 0.2)
      .. controls ++(0.1, 0) and ++(-0.2, 0) ..
      ++(0.2, -0.2);
  \end{scope}

  \node [draw=black, single arrow, single arrow head extend=2mm, minimum height=8mm] at (10, 0) {};
  \node at (10, 0.6) {deform};
  \node at (10, -0.6) {the wall};

  \begin{scope}[shift={(13,0)}]
    \draw[black!30, line width = \h * 1cm] (\X, -\Y) -- (\X, 0) -- (-\X, 0) -- (-\X, \Y);
    \draw[->, >=stealth] (0, -\Y) -- ++(0, \Y * 2) node[above left] {$t$};
    \draw[->, >=stealth] (-\X, 0) -- ++(\X*2, 0) node[below right] {$x$};

    \draw[->, >=stealth, red, semithick] (-0.6, -0.6) -- (-0.3, -0.3);
    \draw[red, semithick] (-0.9, -0.9)
      .. controls ++(0.2, 0) and ++(-0.1, 0) ..
      ++(0.2, 0.2)
      .. controls ++(0.1, 0) and ++(-0.2, 0) ..
      ++(0.2, -0.2);
    \draw[->, >=stealth, blue, semithick] (0.3, 0.3) -- (0.6, 0.6);
    \draw[blue, semithick] (0.6, 0.8)
      .. controls ++(0.2, 0) and ++(-0.1, 0) ..
      ++(0.2, 0.2)
      .. controls ++(0.1, 0) and ++(-0.2, 0) ..
      ++(0.2, -0.2);
  \end{scope}
\end{tikzpicture}
\caption{Exotic scattering as an application of a symmetry operation.\label{fig:illustration}}
\end{figure}
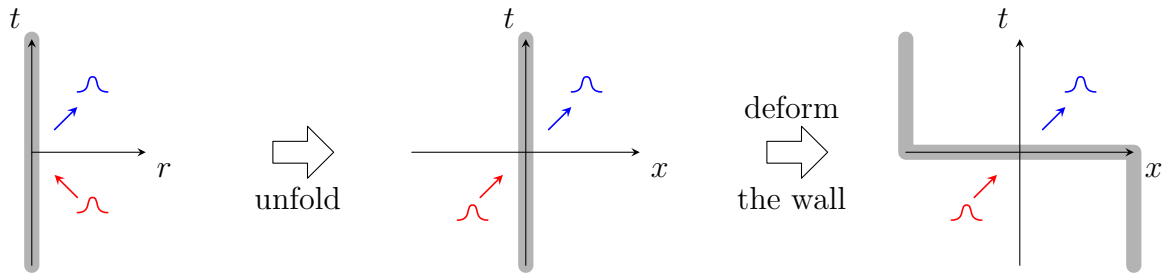

The essential idea is the following.\footnote{%
The authors thank K. Ohmori for enlightening discussions on this point \cite{Ohmori:unpublished}.
}
In the s-wave approximation, we have a two-dimensional setup on a half-line $r>0$, with
both the left-moving and the right-moving degrees of freedom.
Following Polchinski\cite{Polchinski:1984uw}, we unfold the setup and consider the system on the entire real line $-\infty<x<+\infty$ with only the right-moving degrees of freedom,
with a domain wall at $x=0$.
It turns out that this domain wall in the low energy limit is not just conformal but actually topological,
and is of the type corresponding to a (non-invertible) symmetry.
Therefore, the state of the wavepackets after passing through the domain wall
can equally be determined by applying the symmetry to the wavepackets.
An illustration of these ideas is given in Figure~\ref{fig:illustration}.\if0\footnote{%
Therefore, the apparent loss of unitarity in this type of exotic scattering,
if we take the naive point of view that the entirety of the physics can be described within
the Fock space of the elementary fields,
can be said to be related to the non-invertibility of the corresponding symmetry operation.
}\fi

In the rest of the paper, we will carry out a concrete study of this issue
in the unfolded two-dimensional setup where we have four right-moving complex fermions,
with the boundary condition studied by Maldacena and Ludwig \cite{Maldacena:1995pq}.
In Sec.~\ref{sec:circle}, we study the system on a finite circle,
with \emph{two} Maldacena-Ludwig walls inserted. 
We describe the Hilbert space of the system,
and explain how the state of the wavepackets after they pass through a wall can be written down.
In Sec.~\ref{sec:state}, we make a concrete application of the strategy explained in the previous section by writing down the explicit wavefunction of two particles after they pass through the wall. 
We will explicitly see that the energy and the charge are still localized
and the excitation has the expected fractional charge.
We will also study the expectation value of the total number of original particles and anti-particles
in the wavefunction after the scattering. 
We show both analytically and numerically that this number diverges in the limit where the original wavepacket is perfectly localized. 
Finally in Sec.~\ref{sec:conclusions}, we conclude by answering some frequently asked questions,
and discussing future directions.

We have three appendices. 
In Appendix~\ref{app:onewall}, we discuss the system on a circle 
with a \emph{single} Maldacena-Ludwig wall. 
We will see that the Hilbert space can still be written down explicitly, but is not as useful for our purposes.
In Appendix~\ref{app:B}, we discuss the effects of the operators 
of the general form $\exp(\int dx f^a(x) J^a(x))$, where $J^a(x)$ are conserved currents 
and $f^a(x)$ are coefficient functions.
In the main text, we only use the tame case when $f^a(x)$ is nonzero only within a finite interval,
and the topic of this appendix is what happens if we drop this simplifying assumption.
In Appendix~\ref{app:C}, we provide a proof of an analytical estimate needed in Section~\ref{sec:state}.

\section{System on a circle}
\label{sec:circle}
\subsection{Basics of the Maldacena-Ludwig wall}
Let us start by reviewing the boundary condition found by Maldacena and Ludwig. 
We have eight real Majorana-Weyl fermions. Let us say that they transform in the $8_V$ of 
the $so(8)$ symmetry. This symmetry has also two spinor representations, $8_S$ and $8_C$.
The Maldacena-Ludwig boundary condition acts as a linear transformation of $so(8)$ currents,
and exchanges $8_V$ and $8_S$.
This can be thought of as a transformation  by $g\in O(8)_C$ acting on $8_C$
with $\det g=-1$.
For concreteness, we pick the boundary condition specified by 
\begin{equation}
g=\text{diag}(\overbrace{+1,\ldots,+1}^{\text{7 times}},-1)\in O(8)_C,
\label{g}
\end{equation} which preserves $so(7) \subset so(8)$.\footnote{%
This is the boundary condition relevant for the Callan-Rubakov problem.
For the two-channel Kondo problem, the correct choice is 
$$g'=\text{diag}(+1,+1,+1,+1,+1,-1,-1,-1),$$
preserving $so(5)\times so(3)\subset so(8)$.
}
Denoting by $J_{1,2,3,4}$ the currents of $U(1)^4\subset SO(8)_V$,
this $g$ is known to act by the matrix\begin{equation}
g= \frac12 \begin{pmatrix}
+1 & +1 & +1 & +1  \\
+1 & +1 & -1 & -1  \\
+1 & -1 & +1 & -1  \\
+1 & -1 & -1 & +1  
\end{pmatrix}.
\label{gg}
\end{equation}

The same wall also has the following interpretation.
Consider eight Majorana-Weyl fermions $\psi$ in $8_V$.
The $\bZ_2$ action $\psi\mapsto -\psi$  is anomaly-free,
and the gauged theory is still a theory of eight Majorana-Weyl fermions
but transforming in $8_S$.
Therefore, if we perform the $\bZ_2$ gauging on a half-space, 
we have the Maldacena-Ludwig wall on the boundary.
From this description, we can easily derive the fusion rule: \begin{equation}
\text{(Maldacena-Ludwig)}\otimes \text{(Maldacena-Ludwig)}
= \text{(trivial wall)} \oplus \text{($\bZ_2$ wall)},
\label{fusion}
\end{equation}
see e.g.~\cite[Sec.~4.3.1]{Chang:2018iay}.

\subsection{Effect of the walls in the `Heisenberg picture'}
\label{this-subsec}
Let us start by considering the system on a circle, to make the energy spectrum discrete
so that we can have a better grasp of the nature of the states.
We consider the case with two walls, for which the interpretation is far simpler.
We will analyze the case with one wall in Appendix~\ref{app:onewall}.

\begin{figure}[h]
\[
\begin{array}{c@{\qquad\qquad}c@{\qquad\qquad}c}
\vcenter{\hbox{
\begin{tikzpicture}[scale=0.6, line cap=round,line join=round]
  \pgfmathsetmacro{\R}{2}
  \pgfmathsetmacro{\r}{0.5}
  \pgfmathsetmacro{\H}{3}

  \fill[black!20]
    (-\R,0) -- (-\R, \H)
    arc[start angle=180,end angle=110,x radius=\R,y radius=\r]
    -- ++(0, -\H)
    arc[start angle=110,end angle=180,x radius=\R,y radius=\r]
    -- cycle;
  \fill[black!20]
    (-\R,0) -- (-\R, \H)
    arc[start angle=180,end angle=290,x radius=\R,y radius=\r]
    -- ++(0, -\H)
    arc[start angle=290,end angle=180,x radius=\R,y radius=\r]
    -- cycle;

  \draw (0,\H) ellipse[x radius=\R,y radius=\r];
  \draw (0,\H) ellipse[x radius=\R,y radius=\r];
  \draw (-\R,0) -- (-\R,\H);
  \draw (\R,0) -- (\R,\H);
  \draw (-\R,0) arc[start angle=180,end angle=360,x radius=\R,y radius=\r];
  \draw[dashed] (\R,0) arc[start angle=0,end angle=180,x radius=\R,y radius=\r];

  \draw[black!60, very thick]
    ({\R * cos(70)}, {\H - \r * sin(70)}) -- ++(0, -\H);
  \draw[black!60, very thick]
    ({-\R * cos(70)}, {\H + \r * sin(70)}) -- ({-\R * cos(70)}, {\H - \r * sin(70)});
  \draw[black!60, very thick, dashed]
    ({-\R * cos(70)}, {\H - \r * sin(70)}) -- ({-\R * cos(70)}, {\r * sin(70)});

  \draw[->, >=stealth, red, semithick] (0.2, 0.1) -- ++(0.3, 0.3);
  \draw[red, semithick] (-0.1, -0.2)
    .. controls ++(0.2, 0) and ++(-0.1, 0) ..
    ++(0.2, 0.2)
    .. controls ++(0.1, 0) and ++(-0.2, 0) ..
    ++(0.2, -0.2);
  \draw[->, >=stealth, blue, semithick] (0.8, 0.7) -- ++(0.3, 0.3);
  \draw[blue, semithick] (1.2, 1.2)
    .. controls ++(0.2, 0) and ++(-0.1, 0) ..
    ++(0.2, 0.2)
    .. controls ++(0.1, 0) and ++(-0.2, 0) ..
    ++(0.2, -0.2);
\end{tikzpicture}
}} &
\vcenter{\hbox{
\begin{tikzpicture}[scale=0.6, line cap=round,line join=round]
  \pgfmathsetmacro{\R}{2}
  \pgfmathsetmacro{\r}{0.5}
  \pgfmathsetmacro{\H}{3}

  \fill[black!20]
    (-\R,0) -- (-\R, \H * 1.8)
    .. controls (-\R, \H * 1.6) and (-\R * 0.2, \H * 1.6) ..
    (-\R * 0.2, \H * 1.4)
    -- (-\R * 0.2, \H * 0.6)
    .. controls (-\R * 0.2, \H * 0.4) and (\R, \H * 0.4) ..
    (\R, \H * 0.2)
    -- (\R, 0)
    arc[start angle=0,end angle=-180,x radius=\R,y radius=\r]
    -- cycle;
  \fill[black!20]
    (-\R,0) -- (-\R, \H * 1.8)
    .. controls (0, \H * 1.7) and (\R * 0.2, \H * 1.6) ..
    (\R * 0.2, \H * 1.5)
    -- (\R * 0.2, \H * 0.7)
    .. controls (\R * 0.2, \H * 0.4) and (\R, \H * 0.7) ..
    (\R, \H * 0.2)
    -- (\R, 0)
    arc[start angle=0,end angle=-180,x radius=\R,y radius=\r]
    -- cycle;

  \draw (0,\H * 2) ellipse[x radius=\R,y radius=\r];
  \draw (0,\H * 2) ellipse[x radius=\R,y radius=\r];
  \draw (-\R,0) -- (-\R,\H * 2);
  \draw (\R,0) -- (\R,\H * 2);
  \draw (-\R,0) arc[start angle=180,end angle=360,x radius=\R,y radius=\r];
  \draw[dashed] (\R,0) arc[start angle=0,end angle=180,x radius=\R,y radius=\r];
  \draw (-\R,\H) arc[start angle=180,end angle=360,x radius=\R,y radius=\r];
  \draw[dashed] (\R,\H) arc[start angle=0,end angle=180,x radius=\R,y radius=\r];

  \draw[black!60, very thick]
    (-\R, \H * 1.8)
    .. controls (-\R, \H * 1.6) and (-\R * 0.2, \H * 1.6) ..
    (-\R * 0.2, \H * 1.4)
    -- (-\R * 0.2, \H * 0.6)
    .. controls (-\R * 0.2, \H * 0.4) and (\R, \H * 0.4) ..
    (\R, \H * 0.2);
  \draw[black!60, very thick, dashed]
    (-\R, \H * 1.8)
    .. controls (0, \H * 1.7) and (\R * 0.2, \H * 1.6) ..
    (\R * 0.2, \H * 1.5)
    -- (\R * 0.2, \H * 0.7)
    .. controls (\R * 0.2, \H * 0.4) and (\R, \H * 0.7) ..
    (\R, \H * 0.2);

  \draw[red, very thick, decorate, decoration={snake, segment length=\H * 0.05cm, amplitude=0.05cm}] (-\R, \H * 1.8) -- (-\R, \H * 2);
  \draw[red, very thick, decorate, decoration={snake, segment length=\H * 0.05cm, amplitude=0.05cm}] (\R, 0) -- (\R, \H * 0.2);

  \draw[->, >=stealth] (-\R * 1.4, \H * 0.5) node[left] {$\psi(x)$} -- (-\R * 0.8, \H * 0.5);
  \draw[->, >=stealth] (\R * 1.4, \H * 1.5) node[right] {$\widetilde\psi(x)$} -- (\R * 0.8, \H * 1.5);
\end{tikzpicture}
}} &
\vcenter{\hbox{
\begin{tikzpicture}[scale=0.6, line cap=round,line join=round]
  \pgfmathsetmacro{\R}{2}
  \pgfmathsetmacro{\r}{0.5}
  \pgfmathsetmacro{\H}{3}

  \fill[black!20]
    (-\R,0) -- (-\R, \H * 0.5)
    arc[start angle=180,end angle=0,x radius=\R,y radius=\r]
    -- (\R, 0)
    arc[start angle=0,end angle=-180,x radius=\R,y radius=\r]
    -- cycle;

  \draw (0,\H) ellipse[x radius=\R,y radius=\r];
  \draw (0,\H) ellipse[x radius=\R,y radius=\r];
  \draw (-\R,0) -- (-\R,\H);
  \draw (\R,0) -- (\R,\H);
  \draw (-\R,0) arc[start angle=180,end angle=360,x radius=\R,y radius=\r];
  \draw[dashed] (\R,0) arc[start angle=0,end angle=180,x radius=\R,y radius=\r];

  \draw[black!60, very thick, dashed]
    (-\R, \H * 0.5)
    arc[start angle=180,end angle=0,x radius=\R,y radius=\r];
  \draw[black!60, very thick]
    (-\R, \H * 0.5)
    arc[start angle=180,end angle=360,x radius=\R,y radius=\r];

  \draw[red, very thick, decorate, decoration={snake, segment length=\H * 0.05cm, amplitude=0.05cm}] (-\R, 0) -- (-\R, \H);
\end{tikzpicture}
}} \\
\text{(a)} & \text{(b)} & \text{(c)}
\end{array}
\]
\caption{(a) A wavepacket passing through a wall.
(b) We analyze it by merging two walls in two ways. 
The red wavy lines represent the direct sum of an identity wall and a $\bZ_2$ wall.
(c) Its effect can be summarized by the action of the wall wrapping in the spatial direction.
\label{fig:pass}}
\end{figure}
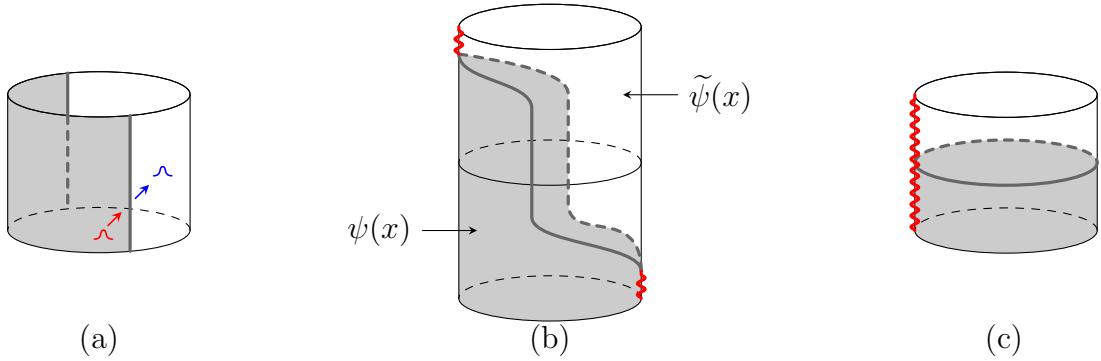

We prepare a localized wavepacket near one wall, and let it pass through the wall
by applying the time-translation operator $e^{iTH}$ for some $T>0$,
see Fig.~\ref{fig:pass} (a).
We analyze this situation by merging two walls in two ways, before and after the wavepacket
goes through the wall, see Fig.~\ref{fig:pass} (b).

Using the fusion rule \eqref{fusion}, the total Hilbert space $\cH_{\text{two walls}}$ of the system at $t\ll 0$
is the sum of the NS sector and the R sector of eight fermions $\psi(x)$ 
on the shaded side of the wall: \begin{equation}
\cH_{\text{two walls}} = \cH_{\psi,\text{NS}}  \oplus \cH_{\psi,\text{R}}.
\label{H-two-walls}
\end{equation}
Let us say that $\psi(x)$ transforms as $8_V$ under the $so(8)_1$ current algebra of the system.
Then, the NS/R sector Hilbert spaces $\cH_{\psi,\text{NS}/\text{R}}$ 
decompose under $so(8)_1$  as \begin{align}
\cH_{\psi,\text{NS}} &= \chi_0 \oplus \chi_V, &
\cH_{\psi,\text{R}} &= \chi_S \oplus \chi_C, 
\end{align} where $\chi_{0,V,S,C}$ are the irreducible representations of
the $so(8)_1$ current algebra whose lowest weight states transform as the identity representation,
$8_V$, $8_S$, $8_C$, respectively.
Similarly, the Hilbert space of the system at $t\gg 0$
is the sum of the NS sector and the R sector of eight fermions $\tilde\psi(x)$ 
on the unshaded side of the wall.
Using the convention that $\tilde \psi(x)$ transforms in $8_S$ of the $so(8)_1$ symmetry,
we similarly have \begin{align}
\cH_{\tilde\psi,\text{NS}} &= \chi_0 \oplus \chi_S, &
\cH_{\tilde\psi,\text{R}} &= \chi_V \oplus \chi_C.
\end{align}

In this description, nothing happens to the state when a wavepacket goes through the wall,
except for acquiring a trivial phase factor by $e^{iHt}$.
For example, 
let us consider a one-particle excitation of $\psi(x)$ in $\chi_V\subset \cH_{\psi,\text{NS}}$.
After passing through the wall, it results in a state in $\chi_V\subset \cH_{\tilde\psi,\text{R}}$.
Similarly, a two-particle excitation of $\psi(x)$ in $\chi_0 \subset \cH_{\psi,\text{NS}}$
becomes a state in $\chi_0\subset \cH_{\tilde\psi,\text{NS}}$.
This is analogous to the `Heisenberg picture' of the time evolution,
in the sense that nothing happens to the state while the operator to be used in its interpretation
changes from $\psi(x)$ to $\tilde\psi(x)$.

Before proceeding, note that the total Hilbert space $
\cH_{\text{two walls}}=\cH_{\psi,\text{NS}}\oplus \cH_{\psi,\text{R}}
=\cH_{\tilde\psi,\text{NS}}\oplus \cH_{\tilde\psi,\text{R}}
$
has actions  of two sets of free fermions fields, $\psi(x)$ and $\tilde \psi(x)$.
The modes of $\psi(x)$ and the modes of $\tilde \psi(x)$ have complicated commutation relations, however, and are not independent of each other.
In particular, the Hilbert space is \emph{not} the tensor product of two Fock spaces
of $\psi(x)$ and $\tilde\psi(x)$.
Rather, $\tilde\psi(x)$ acts in a complicated manner on the Fock space of $\psi(x)$.
It might be of interest to generalize the standard LSZ reduction theorem to such situations.

\subsection{Effect of the walls in the `Schr\"odinger picture'}

For our purposes, it is more useful to go to the `Schr\"odinger picture',
where the operator $\psi(x)$ is kept fixed 
but we apply the transformation $g\in O(8)_C$ given in \eqref{g}
exchanging $\chi_V$ and $\chi_S$ on the total Hilbert space 
$\cH_{\text{tot}}=\cH_{\psi,\text{NS}}\oplus\cH_{\psi,\text{R}}$.
This is an action $U_g$ on the total Hilbert space $\cH_{\text{tot}}$
of the Maldacena-Ludwig wall wrapping the spatial direction,
as shown in Fig.~\ref{fig:pass} (c).\footnote{%
This can be seen as an example of the fact generally proved in a recent paper \cite{Bartsch:2026wqq}
that the action of a non-invertible symmetry operation such as the Maldacena-Ludwig wall
can be represented by a unitary operator, when we consider an appropriate direct sum of 
(un)twisted sectors.
}

The action of $U_g$ in $\chi_0\subset \cH_{\psi,\text{NS}}$
is particularly simple.
Indeed, any state in $\chi_0$ is a linear combination 
of the state of the  form \begin{equation}
J(a_\ell)_{-n_\ell} \cdots J(a_2)_{-n_2} J(a_1)_{-n_1}\ket{0}
\label{Ja}
\end{equation} where $J(a)_{-n}$ is the creation operator of $so(8)_1$
in the direction $a\in so(8)$.
Then $U_g$ simply sends it to \begin{equation}
J(b_\ell)_{-n_\ell} \cdots J(b_2)_{-n_2} J(b_1)_{-n_1}\ket{0}
\label{Jb}
\end{equation} where $b_i= g a_i g^{-1}$. Hereafter, we focus on the NS sector $\mathcal{H}_{\psi,\text{NS}}$.

Our concrete strategy can then be summarized as follows:
\begin{enumerate}
\item Prepare localized wavepackets of $\psi(x)$ excitations 
 in $\chi_0\subset \cH_{\psi,\text{NS}}$.
\item Rewrite it as a linear combination of the states of the form \eqref{Ja}.
\item Apply $U_g$ and obtain a linear combination of the states of the form \eqref{Jb}.
\item Re-interpret the resulting state in terms of $\psi(x)$ excitations.
\end{enumerate}

\begin{figure}
\centering
\begin{tikzpicture}[line cap=round,line join=round]
  \pgfmathsetmacro{\R}{1.5}
  \pgfmathsetmacro{\r}{0.5}

  \draw (0,0) ellipse (\R * 1cm and \r * 1cm);

  \draw[blue] ({\R * cos(240)}, {\r * sin(240)})
    .. controls ++(0.2, 0) and ++(-0.1, 0) ..
    ({\R * cos(250)}, {\r * sin(250) + 0.2})
    .. controls ++(0.1, 0) and ++(-0.2, 0) ..
    ({\R * cos(260)}, {\r * sin(260)});
  \draw ({\R * cos(235)}, {\r * sin(235) - 0.1})
    .. controls ++(-0.1, -0.1) and ++(0, 0.1) ..
    ({\R * cos(250) - 0.1}, {\r * sin(250) - 0.2})
    .. controls ++(0.1, 0.1) and ++(0, -0.1) ..
    ({\R * cos(262)}, {\r * sin(265) - 0.1});
  \node[below] at ({\R * cos(250) - 0.1}, {\r * sin(250) - 0.2}) {$U$};

  \node at (2.2,0) {$\in\chi_V$};

  \begin{scope}[shift={(6,0)}]
    \draw (0,0) ellipse (\R * 1cm and \r * 1cm);
    \draw[blue] ({\R * cos(240)}, {\r * sin(240)})
      .. controls ++(0.2, 0) and ++(-0.1, 0) ..
      ({\R * cos(250)}, {\r * sin(250) + 0.2})
      .. controls ++(0.1, 0) and ++(-0.2, 0) ..
      ({\R * cos(260)}, {\r * sin(260)});
    \draw ({\R * cos(235)}, {\r * sin(235) - 0.1})
      .. controls ++(-0.1, -0.1) and ++(0, 0.1) ..
      ({\R * cos(250) - 0.1}, {\r * sin(250) - 0.2})
      .. controls ++(0.1, 0.1) and ++(0, -0.1) ..
      ({\R * cos(262)}, {\r * sin(265) - 0.1});
    \node[below] at ({\R * cos(250) - 0.1}, {\r * sin(250) - 0.2}) {$U$};

    \draw[blue] ({\R * cos(80)}, {\r * sin(80)})
      .. controls ++(0.2, 0) and ++(-0.1, 0) ..
      ({\R * cos(70)}, {\r * sin(70) + 0.2})
      .. controls ++(0.1, 0) and ++(-0.2, 0) ..
      ({\R * cos(60)}, {\r * sin(60)});
    \draw ({\R * cos(82)}, {\r * sin(82) + 0.3})
      .. controls ++(0.1, 0.1) and ++(-0.1, -0.1) ..
      ({\R * cos(70) + 0.05}, {\r * sin(70) + 0.4})
      .. controls ++(0, -0.1) and ++(0, 0.1) ..
      ({\R * cos(55)}, {\r * sin(55) + 0.3});
    \node[above] at ({\R * cos(70) + 0.2}, {\r * sin(70) + 0.4}) {$U'$};

    \node at (2.2,0) {$\in\chi_0$};
  \end{scope}
\end{tikzpicture}
\caption{Localized excitations and the corresponding states.
Which of $\chi_{0,V,S,C}$ a particular state is in depends on 
the entire configuration of localized excitations,
and is not an intrinsic property of an excitation localized in a region $U$.  \label{fig:sectors}}
\end{figure}
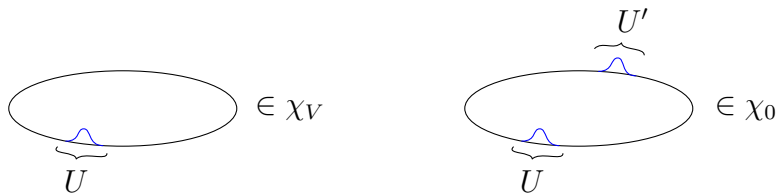

There are two comments.
The first comment concerns 
the fact  that not all the states are in $\chi_0\subset\cH_{\psi,\text{NS}}$.
For example, a single fermion excitation is in $\chi_V$ and not in $\chi_0$.
But as long as we are interested in the properties of  excitations localized in a region $U$,
we can always add some other excitations with opposite charges 
localized in a  region $U'$ far away from $U$,
so that the total state is in $\chi_0$.
In a relativistic local quantum field theory,
additional excitations in $U'$ should not affect the physics of excitations in $U$.\footnote{%
\label{nonsensical-philosophy}
For example, suppose that the circumference of the circle is about the distance 
between us and the Andromeda galaxy.
Suppose furthermore that we prepare a $\psi$ wavepacket
close to one of the walls, separated only by about the size of a human being.
Then, what happens to this $\psi$ wavepacket when it goes through the wall
should not be affected by the presence of an additional $\psi$ wavepacket 
in a galaxy far, far away.
The sector theory in algebraic quantum field theory is based on this idea,
see e.g.~the textbook \cite{Haag:1996hvx}.
}
In this sense, which of the four subspaces $\chi_{0,V,S,C}$ a particular state is in
is not an intrinsic property of an excitation localized in $U$.
This means that studying states in the $\chi_0$ sector should be enough for our purposes.
See Fig.~\ref{fig:sectors} for an illustration.

The second comment is that our procedure, summarized as the four steps stated above,
is simply the known procedure consisting of
(i) preparing fermions,
(ii) bosonizing the fermions,
(iii) rotating the bosons,
(iv) and then re-fermionizing the bosons.
In this sense, what we did so far is no different from 
what was already used in the original works in the late 80s and the early 90s.
For example, it was used in the paper \cite{Maldacena:1995pq} by Maldacena and Ludwig,
which we heavily depend on.
Hopefully, the discussion presented here sheds somewhat new light 
on the physics of this scattering, such as its relation to walls
implementing non-invertible symmetry actions,
and the classes of states for which this procedure is safely applicable.

\section{Wavefunction of two exotic particles}
\label{sec:state}

Let us now carry out the strategy explained in the last section, and determine the wavefunction
of the wavepackets of two particles after they pass through the wall, in terms of the original variables.
For states spread out on the entire spatial $S^1$, this is very easy.
For example, consider the two-particle state $\psi^1_{-1/2} \bar\psi^1_{-1/2}\ket 0$.
Here and below, we use four chiral complex fermions $\psi^{i}(x)$, $\bar\psi^i(x)$, where $i=1,2,3,4$.
This state is simply equal to $J(a)_{-1}\ket 0$ for some $a\in so(8)$.
Then \begin{equation}
U_g\psi^1_{-1/2} \bar\psi^1_{-1/2}\ket 0 = U_g J(a)_{-1}\ket 0= J(b)_{-1}\ket 0,
\end{equation} where $b=g a g^{-1}$. The right hand side can then be rewritten 
in terms of linear combinations of $\psi^{i}_{-1/2} \bar\psi^{j}_{-1/2}\ket 0$.

We are, however, more interested in the fate of the localized wavepackets. 
For this, we would like to study $U_g$ applied on $\psi^1(x_1)\bar\psi^1(x_2)\ket 0$, say.
Using bosonization, we can write\footnote{We implicitly understand the right hand side to be normal-ordered, so its overlap with the vacuum is 1. The left hand side is normalized so that the same is true there as well.} \begin{equation}
\frac{\psi^1(x_1)\bar\psi^1(x_2)\ket 0}{\bra0\psi^1(x_1)\bar\psi^1(x_2)\ket 0}=\exp\left( 2\pi i\int dx\,  c(x) J^1(x) \right) \ket 0,
\label{before}
\end{equation}
where $J^i(x)=\psi^i(x) \bar\psi^i(x)$ and \begin{equation}
c(x)=\begin{cases}
1, & (x_1< x< x_2), \\
0, & \text{otherwise}.
\end{cases}
\end{equation} 
Using the explicit form of $g$ given in \eqref{gg}, we then have \begin{equation}
U_g \left(\frac{\psi^1(x_1)\bar\psi^1(x_2)\ket 0}{\bra0\psi^1(x_1)\bar\psi^1(x_2)\ket 0}\right)=\exp\left(2\pi i \int dx\,  c(x) \frac12 (J^1(x)+J^2(x)+J^3(x)+J^4(x)) \right) \ket 0.
\label{after}
\end{equation}
We are done, then, if we can rewrite states of the form \begin{equation}
\exp\left(\int dx f(x) J(x)\right)\ket 0 \label{general-state}
\end{equation} 
for general functions $f(x)$ in terms of the fermion creation operators.\footnote{%
For simplicity, we only consider the case when $f(x)$ is a periodic function on $S^1$.
Then, the state \eqref{general-state} clearly exists in the Hilbert space of theory in question,
as we can interpolate it from the vacuum by considering the one-parameter family 
$\exp\left(\alpha\int dx\, f(x) J(x)\right)\ket 0$ of states  for $\alpha\in [0,1]$, say.
The status of states of this form with less restrictions on the functions $f(x)$ will be discussed
in Appendix~\ref{app:B}.
}

\subsection{Computation of the transformed wavefunction}
\label{sec:computation}
For ease of computation, we move to the radial quantization where the spatial $S^1$ is on the unit circle $|z|=1$,
for which we use the coordinate $\theta$ via the relation $z=e^{i\theta}$.
We consider only a single complex chiral fermion $\psi(z)$ and drop the superscript distinguishing the four flavors.
We use the standard normalization so that $\bar\psi(z)\psi(w) \sim 1/(z-w)$.
We then define $J(z)=\bar\psi(z)\psi(z)$, and consider the state given by \begin{equation}
e^{i\alpha A_0}\ket 0, \qquad \text{where}\qquad A_0=\oint \frac{dz}{2\pi i} f(z)J(z).
\end{equation}
Here we separated an overall coefficient $\alpha$ for later convenience, 
although we could have included it in $f(z)$ by rescaling it by $\alpha$.
We need the case $\alpha=1$ and $\alpha=1/2$ to study the wavepackets before and after the transformation, respectively.

It is standard to normal-order the exponential with respect to the modes of $J(z)$, so that we have 
\begin{equation}
e^{i\alpha A_0}\ket 0 = e^{i\alpha A_{}} \ket 0
\end{equation} 
up to an overall normalization, where 
\begin{equation}
A_{}=\oint \frac{dz}{2\pi i} f_{<}(z)J_{\geq}(z).\label{AfJ}
\end{equation}
Here we define, for any function $g(z)\equiv \sum_n g_{-n}z^n$, $g_{<}(z)\equiv \sum_{n<0} g_{-n}z^n$ and $g_{\geq}(z)\equiv \sum_{n\textcolor{black}{\geq} 0} g_{-n}z^n$.
For operators, those subscripted with $\geq$ are the creation operators which we call ``positive'', 
and likewise for $<$.
For fermions we use
\begin{align}
    \psi_{\geq}(z)\equiv \sum_{n\geq 0} \psi_{-n-1/2}z^n
\end{align} and similarly for $\bar \psi$. For the current we use
\begin{align}
    J_{\geq}(z)=\sum_{n\ge0}J_{-n-1}z^n.
\end{align}

We are, however, interested in performing the normal ordering with respect to the modes of $\psi(z)$ and $\bar\psi(z)$.
It is straightforward to convince oneself that the result is also a Gaussian, so we have \begin{equation}
e^{i\alpha A_{}}\ket 0 = e^{B^{(\alpha)}} \ket 0, \quad \text{where}\quad B^{(\alpha)}=\oint \frac{dw_1}{2\pi i} \oint \frac{dw_2}{2\pi i}D^{(\alpha)}(w_1,w_2)\bar\psi_{\geq}(w_1)\psi_{\geq}(w_2)
\end{equation} for some function $D^{(\alpha)}(w_1,w_2)$.
This is an equality rather than a proportionality, because both $e^{i\alpha A}\ket{0}$ and $e^{B^{(\alpha)}}$ contain $\ket{0}$ with unit coefficient.

Now, to determine $D^{(\alpha)}(w_1,w_2)$, we consider the equality 
\begin{align}
    \braket{0|\bar{\psi}(z_1)\psi(z_2)e^{i\alpha A}|0}
    =
    \braket{0|\bar{\psi}(z_1)\psi(z_2)e^{B^{(\alpha)}}|0}.
\end{align}
Using $e^{-i\alpha A_{}}\psi(\theta)e^{i\alpha A_{}}=e^{-2\pi i\alpha f_{<}(\theta)}\psi(\theta)$, the left hand side
can be evaluated as 
\begin{align}
    \mathrm{(LHS)}
    &=\braket{0|e^{-i\alpha A_{}}\bar{\psi}(z_1)\psi(z_2)e^{i\alpha A_{}}|0}\\
    &=e^{2\pi i\alpha (f_{<}(z_1)-f_{<}(z_2))}
    \braket{0|\bar{\psi}(z_1)\psi(z_2)|0}.
\end{align}
On the other hand, the RHS becomes
\begin{align}
    \mathrm{(RHS)}
    &=
    \braket{0|\bar{\psi}(z_1)\psi(z_2)|0}+
    \oint\frac{dw_1}{2\pi i} \oint \frac{dw_2}{2\pi i}\frac{D^{(\alpha)}(w_1,w_2)}{(z_1-w_2)(z_2-w_1)}\\
    &=\braket{0|\bar{\psi}(z_1)\psi(z_2)|0}+
    D^{(\alpha)}_{<,<}(z_2,z_1). \label{RHS}
\end{align}

Swapping $z_{1}$ and $z_2$ with respect to \eqref{RHS}, we obtain
\begin{align}
    D^{(\alpha)}_{<,<}(z_1,z_2)=\frac{1}{z_2-z_1}(e^{2\pi i\alpha (f_{<}(z_2)-f_{<}(z_1))}-1).
    \label{Ddef}
\end{align}
Note that it has a finite limit when $z_2\to z_1$: The $-1$ in the parentheses removes the singularity at $z_1= z_2$. If one adopts a prescription of deforming the integration contour so that $|z_1|<|z_2|$, the $-1$ can be dropped.

We choose $f(\theta)$ to be a step function $c(\theta)$ from $\theta_1$ to $\theta_2$.
Written in terms of the holomorphic coordinate and setting $\zeta_{1,2}=e^{i\theta_{1,2}}$, we have
\begin{align}
    2\pi if_{<}(z)=\log\left(\frac{z-\zeta_2}{z-\zeta_1}\right),
\end{align}
where it is convenient to set 
\begin{equation}
\zeta_{1,2}=  e^{-\epsilon+i\theta_{1,2}} \label{zetadef}
\end{equation}
for a small positive real $\epsilon$, so that $|\zeta_{1,2}| < |z|=1$
and later take the limit $\epsilon\to 0$.

As a consistency check, we can put $\alpha=1$ and then we see that 
\begin{align}
B^{(\alpha=1)}&= \oint\frac{dw_1}{2\pi i} \oint \frac{dw_2}{2\pi i}\frac{1}{w_2-w_1} 
\left(\frac{w_1-\zeta_1}{w_1-\zeta_2}\frac{w_2-\zeta_2}{w_2-\zeta_1} -1\right)
\bar\psi_\ge(w_1) \psi_\ge(w_2) \\
&=(\zeta_2-\zeta_1)\bar{\psi}_\geq(\zeta_1)\psi_\geq(\zeta_2).
\end{align} Then \begin{equation}
e^{B^{(\alpha=1)}}\ket 0 = \ket0 +(\zeta_2-\zeta_1)\bar{\psi}_\geq(\zeta_1)\psi_\geq(\zeta_2)\ket0
\label{regularized}
\end{equation} without any further terms.
This is as expected.
In contrast, when $\alpha=1/2$, we have \begin{equation}
B^{(\alpha=\frac12)}=\oint\frac{dw_1}{2\pi i} \oint \frac{dw_2}{2\pi i} \frac{1}{w_2-w_1}
\left(\sqrt{\frac{w_1-\zeta_1}{w_1-\zeta_2}\frac{w_2-\zeta_2}{w_2-\zeta_1}}-1\right) 
\bar\psi_\ge(w_1) \psi_\ge(w_2)
\label{final-result}
\end{equation} 
without any further immediate simplifications. 


In the following, it is actually useful to consider the wavefunctions defined for small nonzero $\epsilon>0$
in \eqref{zetadef},
since it makes the wavepackets not exactly point-localized, making the norms of wavefunctions finite.
This corresponds to wavepackets localized around $z_{1,2}=e^{i\theta_{1,2}}$ with width $\epsilon$.
We can then study what happens in the limit $\epsilon\to 0$.

\subsection{Properties of the transformed wavefunction}


We succeeded in obtaining the explicit wavefunction for the wavepackets after the application of the transformation $U_g$.
Let us now compute some of its physical properties.

\subsubsection{Energy density and charge density}
\label{sec:local}
We will start by considering the expectation values of the energy density $\braket{T(z)}$ and the charge density $\braket{J^a(z)}$,
evaluated in the state $\ket\Psi :=U_g \psi^1(z_1) \bar\psi^1(z_2)\ket 0$.
To be more explicit, we might want to consider a regularized version $\ket{\Psi(\epsilon)}$
where the wavepackets are slightly smeared.

Actually, we do not have to use the explicit expression \eqref{final-result} for this state
to evaluate these expectation values,
since the energy density is invariant, $U_g^{-1} T(z) U_g=T(z)$,
and $U_g^{-1} J^a(z)U_g$ is simply given by the action of $g$ on the superscript $a$.

This leads to the following somewhat obvious answer:
in the transformed state $\ket{\Psi(\epsilon)}$, the energy density and the charge density 
are still localized at $z=z_1$ and $z=z_2$.
Furthermore,  the charge density is such that its local integral around $z\sim z_1$ or $z\sim z_2$
is given by half integers, $(Q^1,Q^2,Q^3,Q^4) = (1/2,\,1/2,\,1/2,\,1/2)$.
These local excitations travel to the right at the speed of light.

The only additional information we gained from having an expression \eqref{final-result} is that 
the wavefunction satisfying these features \emph{does exist} within  the standard free Fock space
constructed out of fermions $\psi^i$ and $\bar\psi^i$, which only have integer charges.\footnote{%
This fact might have been perfectly clear for some of the readers, 
but this was a reassuring observation to the authors of this paper.}

\subsubsection{The average number of (anti)-particles}
\label{sec:number}
Next, we would like to consider the sum of the number of particles \emph{and} anti-particles, $\braket{N}$, contained in the transformed state $\ket{\Psi}$.\footnote{%
Here, we again focus on one flavor and drop the superscript of complex chiral fermions $\psi(z)$.
The total number for the four-flavor case is thus $4N$.}
Note that the difference of the number $N_+$ of the particles and the number $N_-$ of anti-particles is a conserved charge
and therefore is an integral of a local quantity,
but the sum $N=N_+ + N_-$ is not.
Therefore, it is a quantity inherently tied to the structure of the Fock space
as defined using the mode expansion, and is much harder to compute.
Our aim here is to show that $\braket{N}$ diverges as we take the limit $\epsilon\to 0$
of perfectly localized wavepackets.
We first estimate the divergence theoretically,
and then provide a numerical check.

\paragraph{Theoretical analysis:}
Let us write the state as \begin{equation}
\ket{\Psi}=\exp\left(\sum_{n,m\ge 1} D_{nm} \bar\psi_{-n+1/2} \psi_{-m+1/2}\right)\ket0.
\end{equation}
Regarding $D_{nm}$ as a matrix and denoting their singular values
(i.e.~the positive square roots of eigenvalues of $D^\dagger D$)
as $s_1\ge s_2\ge \cdots \ge 0$,
it is immediate that the logarithm of the norm and the average number of particles of the state $\ket{\Psi}$ are given by
\begin{equation}
    \log\braket{\Psi|\Psi}=\sum_{n=1}^\infty\log(1+s_n^2),
    \qquad\braket{N}=2\sum_{n=1}^\infty\frac{s_n^2}{1+s_n^2}.
    \label{formula}
\end{equation}
We use these relations to give an upper bound and a lower bound for $\braket{N}$
in terms of $\log\braket{\Psi|\Psi}$. 

For this purpose, note that 
\begin{equation}
    \frac{x}{1+x}\le\log(1+x)
\end{equation} and also that the function 
\begin{equation}
    g(x)=\frac{x}{1+x} (\log(1+x))^{-1}
\end{equation}
is decreasing for $x\ge 0$.
Therefore,
\begin{equation}
    \braket{N}=2\sum_{n=1}^\infty\frac{s_n^2}{1+s_n^2}\le2\sum_{n=1}^\infty\log(1+s_n^2)=2\log\braket{\Psi|\Psi}
\end{equation}
and 
\begin{equation}
    \braket{N}=2\sum_{n=1}^\infty\frac{s_n^2}{1+s_n^2}\ge 2 g(s_1^2)\sum_{n=1}^\infty\log(1+s_n^2)=2g(s_1^2)\log\braket{\Psi|\Psi}.
    \label{lowerbound}
\end{equation}
We therefore need the information on the norm $\braket{\Psi|\Psi}$
and the largest singular value $s_1$ of $D_{mn}$.

It is straightforward to compute the norm  $\braket{\Psi|\Psi}$.
We go back to the expression $\ket{\Psi}=e^{i A/2}\ket0$,
where $A$ was given in \eqref{AfJ} and note that $\alpha=1/2$ for the transformed state $\ket{\Psi}$.
Then \begin{equation}
\braket{\Psi|\Psi} 
= \braket{0| e^{-iA^\dagger/2} e^{iA/2}|0 }
= e^{[A^\dagger,A]/4}.
\end{equation} It is routine to compute $[A^\dagger,A]$, and we find \begin{equation}
\braket{\Psi|\Psi} =\left|\frac{1-e^{-2\epsilon+i(\theta_2-\theta_1)}}{1-e^{-2\epsilon}}\right|^{1/2}.
\end{equation} Therefore, \begin{equation}
\log\braket{\Psi|\Psi} = \frac{1}{2}\log\left|\frac{1-e^{-2\epsilon+i(\theta_2-\theta_1)}}{1-e^{-2\epsilon}}\right|.
\end{equation}

It is more tricky to estimate the largest singular value $s_1$.
Note that we need to bound $g(s_1^2)$ from below in \eqref{lowerbound}.
As $g(x)$ is a decreasing function, this means that we need to bound $s_1$ from above.
One way to do so is as follows.
Note that
\begin{equation}
s_1^2 \le \sum_{i=1}^\infty s_i^2 = \mathop{\mathrm{tr}} D^\dagger D= \int_0^{2\pi}\frac{d\phi_1}{2\pi}\int_0^{2\pi}\frac{d\phi_2}{2\pi}|D_{<,<}(e^{i\phi_1},e^{i\phi_2})|^2.
\end{equation}
In Appendix~\ref{app:C}, we analyze the right hand side and show that it is bounded from above by
$C(\log\frac1\epsilon)^2$, where $C$ is an $\epsilon$-independent constant.
Then $g(s_1^2)$ is bounded from below by $\sim (\log\log\frac1\epsilon)^{-1}$.

Combining the information, we conclude that 
\begin{equation}
    O\left(\frac{\log\frac{1}{\epsilon}}{\log\log\frac{1}{\epsilon}}\right)\lesssim\braket{N}\le\log\left|\frac{1-e^{-2\epsilon+i(\theta_2-\theta_1)}}{1-e^{-2\epsilon}}\right|\sim O\left(\log\frac{|\theta_1-\theta_2|}{\epsilon}\right).
\end{equation}
In particular, it diverges as $\epsilon \to 0$, i.e.~as the wavepackets are more localized.

\paragraph{Numerical analysis:}

We can also compute $\braket{N}$ numerically and show that $\braket{N}=O(\log \frac{1}{\epsilon})$, \emph{i.e.}, that it saturates the upper bound given analytically.
This can be done by numerically evaluating the eigenvalues of the linear operator $K\equiv D^\dagger D$, which can be given as an integral kernel
\begin{equation}
    \begin{gathered}
        K(z_2,z_1)\equiv \oint \frac{dw_1}{(2\pi i)^2}\,\left[S(z_1)^{\frac{1}{2}}\frac{\sqrt{\frac{(w_1-\zeta_1^{[+]})(w_1-\zeta_1^{[-]})}{(w_1-\zeta_2^{[+]})(w_1-\zeta_2^{[-]})}}}{(z_1-w_1)(z_2-w_1)}S(z_2)^{\frac{1}{2}}\right],\\
        \text{where}\quad
        S(z)\equiv \sqrt{\frac{(z-\zeta_2^{[+]})(z-\zeta_2^{[-]})}{(z-\zeta_1^{[+]})(z-\zeta_1^{[-]})}},
    \end{gathered}
\end{equation}
where $\zeta_{1,2}^{[\pm]}\equiv e^{\pm \epsilon}\zeta_{1,2}$ reflecting the fact that we have regulated $D$ by moving the operator insertion slightly inside the unit circle, and consequently slightly outside for $D^\dagger$.
The contour for the $w_1$ integral is as always the unit circle so it encloses two of the branch points.
The branches of square roots will be chosen so that $K$ becomes a positive operator as per physics.

We now perform a conformal transformation that maps $(\zeta_{2}^{[-]},\,\zeta_{1}^{[-]},\,\zeta_{1}^{[+]},\,\zeta_{2}^{[+]})\mapsto (0,\,1,\,\infty,\, \eta)$.
We can use the translational symmetry to set $\zeta_{1}=e^{-i\theta/2}$ and $\zeta_{2}=e^{i\theta/2}$ (so $\theta_1 = -\theta/2$ and $\theta_2 = \theta/2$), and
then the cross-ratio $\eta$ becomes a negative real number,
\begin{align}
    \eta = -\frac{\sinh^2 \epsilon}{\sin^2(\theta/2)} = O(\epsilon^2).
\end{align}
We can then rewrite the kernel as
\begin{align}
    K(z_1,z_2)=-\frac{1}{4\pi^2}\left({\frac{z_1(z_1-\eta)}{z_1-1}}\right)^{\frac{1}{4}}\frac{F(z_1)-F(z_2)}{z_1-z_2}\left({\frac{z_2(z_2-\eta)}{z_2-1}}\right)^{\frac{1}{4}},
\end{align}
where
\begin{align}
    F(z)\equiv \int_{-\infty}^\eta dw_1\, \left[\frac{1}{w_1-z}\sqrt{\frac{w_1-1}{w_1(w_1-\eta)}}\right]
    =\frac{2i}{z (-\eta)^{1/2}} \left((z-1) \mathop{\mathtt{\Pi}}\left[\frac{z}{\eta},\frac{1}{\eta}\right]+\mathop{\mathtt{K}}\left[\frac{1}{\eta}\right]\right).
\end{align}
The functions $\mathtt{K}$ and $\mathtt{\Pi}$ denote  complete elliptic integrals defined as
\begin{align}
    \mathtt{K}[k^2]\equiv \int_0^1 \frac{dt}{\sqrt{(1-t^2)(1-k^2t^2)}}, \quad \mathtt{\Pi}[\alpha^2,k^2]\equiv \int_0^1 \frac{dt}{\sqrt{(1-t^2)(1-k^2t^2)}(1-\alpha^2t^2)}.
\end{align}
Note a slightly unconventional notation for the variables representing the moduli.

By using this expression, one can straightforwardly discretise the kernel on the coordinate space.
We can then compute all the eigenvalues of discretised $K$ at finite and fixed $\eta$.
We denote them as $\kappa_n$ from $n=1$ to $n=d$ when the discretised matrix is a $d\times d$ matrix.
We also sort them from the largest to the smallest as in the case of $s^2_n$, which we expect $\kappa_n$ to approximate.

As for our discretisation scheme,
we first mapped $z$ to a new coordinate using the exponential function to resolve sudden changes in $K(z_1,z_2)$ near $z_{1,2}=0$.
We then used the Gauss-Jacobi quadrature to resolve the singularity as $z_{1,2}\to 1$.
However, we also found that the final result is almost independent of discretisation schemes we have chosen, which is reassuring.
We also checked that the result is stable against changing the size $d$ of the finite matrix that approximates the kernel.
We picked $d=1000$ to display the following result.

\begin{figure}
    \centering
    \includegraphics[width=0.7\textwidth]{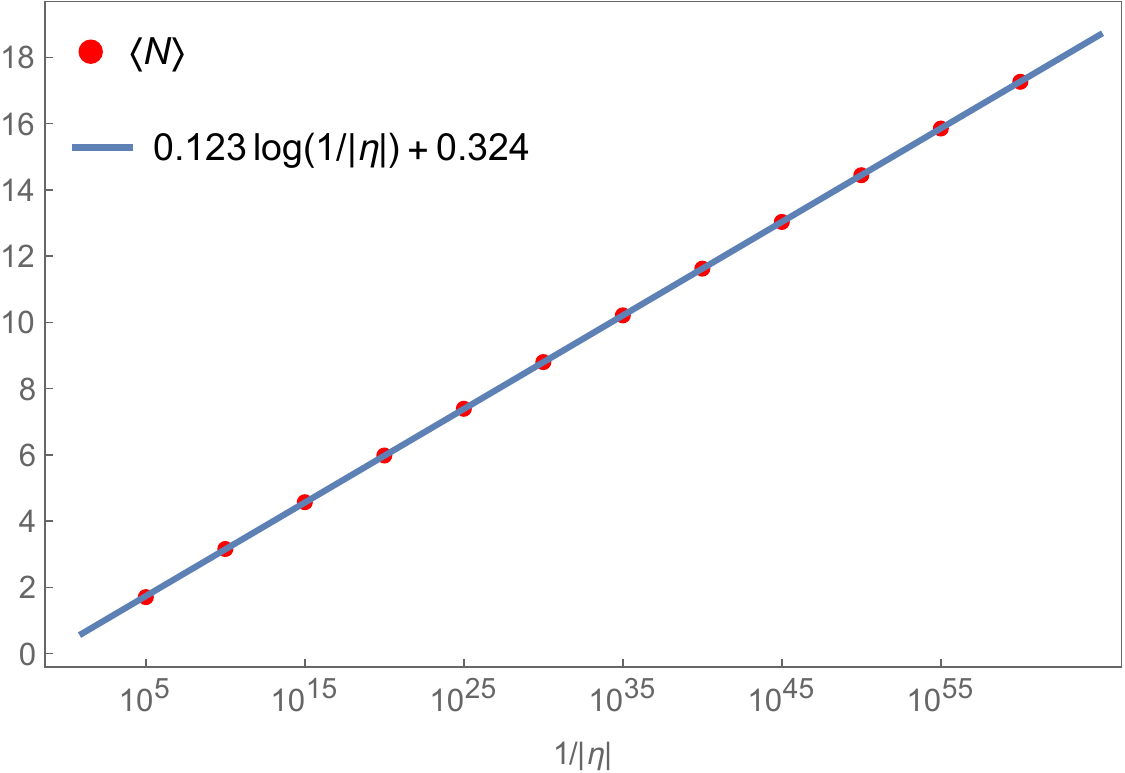}
    \caption{Numerical expectation values of particle number in the state $\ket{\Psi}$ at various values of $\abs{\eta}=O(\epsilon^2)$, with a linear fit in $\log (1/\abs{\eta})$ for comparison.}
    \label{fig:aaa}
\end{figure}

After all the procedures above, we can compute the average number of particles by using the formula
\begin{align}
    \braket{N}=2\sum_{n=1}^{d}\frac{\kappa_n}{1+\kappa_n}.
\end{align}
We plot it at various negative and small $\eta$ as shown in Figure \ref{fig:aaa}.
We can clearly see that $\braket{N}$ scales as $O(\log \frac{1}{\abs{\eta}})=O(\log \frac{1}{\epsilon})$ as promised -- it is strikingly well fitted with\footnote{The leading coefficient is very close to $1/8$, which is the operator dimension of $e^{i\phi/2}$. It would be interesting to offer an analytic explanation making use of the fact that our integral is a generalised version of the Coulomb-gas integral.}
\begin{align}
    \braket{N}\approx 0.123\log\left(\frac{1}{\abs{\eta}}\right)+0.324 = O(\log \epsilon^{-1}).
\end{align}

\section{Conclusions and Outlook}
\label{sec:conclusions}

In this paper, we analyzed the state of the wavepackets of a fermion and an anti-fermion
in a two-dimensional massless system, after they are reflected by the Maldacena-Ludwig boundary
which transforms them into exotic excitations with fractional charges. 
We performed this analysis by first unfolding the system, 
and deformed the resulting Maldacena-Ludwig wall using the fact that it was actually topological.
We then explicitly wrote down the wavefunction after the excitations passed through the wall,
and studied some of its properties.
We found that the energy and the charge of the excitations are indeed localized,
and that it has the expected fractional charge. 
We also computed the expectation value of the number of original fermions and anti-fermions,
and found it to diverge when we make the excitation to be perfectly localized. 

We would like to conclude this paper by answering questions some of the readers might have. 
The list of the questions here is based on the ones 
the authors asked themselves while working on this paper, 
and also on the ones the authors received when they gave the talk on the contents of the paper.

\newcounter{qa}
\def\question#1{\refstepcounter{qa}
\begin{mdframed}[linecolor=black!0,backgroundcolor=black!5]
\noindent \textbf{Q\theqa.}\quad #1
\end{mdframed}
}
\long\def\answer#1{
\begin{mdframed}[linecolor=black!0,backgroundcolor=black!0]
\noindent \textbf{A\theqa.}\quad #1
\end{mdframed}
}

\question{How is it possible to have fractionally charged excitations in the ordinary Fock space,
when it clearly contains only integer-charged states?}

\answer{One needs to distinguish the total charge $\int_{-\infty}^{\infty} dx\, J(x)$ 
and the local integrated charge density $\int_{-\epsilon}^\epsilon dx\, J(x)$  around a single wavepacket.
The total charge is indeed integral, while the local integrated charge density 
has the expected fractional value, as we discussed in Sec.~\ref{sec:local}.
}

\question{With two excitations, the total charge is an integer, for which there was no question about the existence of the state.
Did we actually learn anything new in this analysis?}

\answer{The explicit form of the state with two exotic excitations, \eqref{final-result} given in Sec.~\ref{sec:computation}, might be of some interest.
We can also learn a couple of things about a single excitation on an infinite spatial line
by taking a suitable limit.

For example, the expectation value $\braket{N}$ of the combined number of original fermions and anti-fermions  
on the unit circle
was found to behave as $O(\log \frac{\theta}{\epsilon})$ where
$\theta$ was the separation between two excitations
and $\epsilon$ was the localized width of each of the two excitations.
On a circle of radius $R$, this becomes $O(\log \frac{L}{W})$,
where $L\equiv R\theta$ is the physical separation of two excitations and $W\equiv R\epsilon$ is the width of each wavepacket.
This has a finite limit when we take $L$ and $W$ fixed and take $R\to \infty$.
To separate the two excitations, we can further take $L\to \infty$ while keeping $W$ finite.
We find that the expectation value $\braket{N}$ of a \emph{single} exotic excitation
would then be infinite, even before taking the strictly localized limit $W\to 0$,
in contrast to the case with two excitations.
}

\question{It was repeatedly mentioned that the expectation number $\braket{N}$ of the original fermions and anti-fermions diverges in the wavefunction after the scattering,
when we try to localize the wavepackets. Does this mean that the state is sick?}

\answer{\label{Q:sick}
If we insist on using the original fermion fields $\psi$ to describe the scattered wavefunction, 
indeed there are some infinities associated to it.
However, as long as we only study the properties measurable by the current $J$
and the energy-momentum tensor $T$, the state after the scattering is as good as 
the state before the scattering.

It should also be mentioned that the situation is mutual, in the sense that 
there are new fermion fields $\tilde\psi$ in terms of which 
the scattered wavefunction is simply a two-particle state.
Then, it is the original wavefunction before the scattering that has a divergent number of 
excitations of $\tilde\psi$.
Therefore, the answer to this question depends on the perspective one takes.
}

\question{The system on a circle with two Maldacena-Ludwig walls was used in the main text.
Is it possible to consider the system on a circle with one Maldacena-Ludwig wall, instead?}

\answer{The Hilbert space can be explicitly described, as we give more details 
in Appendix~\ref{app:onewall}.
It turns out that not only the exotic excitation after the scattering
but also the original fermionic excitation before the scattering are hard to be
explicitly given in this description.
Therefore, the circle with a single Maldacena-Ludwig wall is not very useful for our purposes.

The authors' opinion about the number of the Maldacena-Ludwig walls 
on a circle is that it should not matter,
since we can always make the circle very,  very large and 
concentrate on the physics local to a single wall.
As such, the authors do not think there is any philosophical merit to have exactly one wall on a circle.
}

\question{It is sometimes said that the scattered particle corresponds to an operator 
attached at the end of a topological defect. What is the relation between this description 
and the discussion in this paper?}

\answer{\label{Q:defect}
Our transformed state has the general form $\exp(i\int dx f^a(x) J^a(x)) \ket0$,
where $J^a$ are $so(8)$ currents and $f^a$ are coefficient functions.
Therefore it can also be considered as a position-dependent symmetry transformation. 
With one exotic particle at $x_1$ and its anti-particle at $x_2$, 
$f^a(x)$ starts from $0$ when $x\ll x_1$,
takes some constant value $f^a:=f^a(x)$ when $x_1 < x< x_2$,
and comes back to $0$ again when $x_2 \ll x$.
Therefore, we can think of this setup as applying a symmetry transformation 
$g:=e^{if^a T^a} \in SO(8)$ only between $x_1$ and $x_2$.
As such, this state can be said to be attached to a topological defect specified by $g$, 
stretching between $x_1$ and $x_2$.
We discuss slightly more about this issue in Appendix~\ref{app:B}.
  }

\question{Is there anything we can learn about the issues surrounding the Callan-Rubakov effect in four dimensions from the analysis performed here?}

\answer{\label{Q:4d}
One thing we can say is that the explicit wavefunction \eqref{final-result}
after two particles get scattered is still a valid wavefunction in the four-dimensional setting
if we regard the operators $\psi$ and $\bar\psi$ in \eqref{final-result}
as the s-wave part of the four-dimensional fundamental fermions. 
What properties could actually be obtained from this explicit four-dimensional expression
is a distinct but interesting question.
}

\question{The four-dimensional wavefunction suggested in \textbf{A\ref{Q:4d}} 
is spherically symmetric, and
would be `attached' in the sense of \textbf{A\ref{Q:defect}}
to a volume-defect filling the region between two spherical shells.
What happens when we measure this wavefunction?
Do we get a spherically-symmetric result, or a result localized in the angular direction?}

\answer{The authors are not sure, but the following discussion might be of some use.
Even in the case of the scattering of a  fermion by an ordinary point-like target at $(0,\,0,\,0)$, 
the scattered wavefunction can purely be in the s-wave and completely spherical.  
That said, if we make two successive localized measurements of the scattered wavefunction, 
one at $(x_1,\,y_1,\,z_1)$ and another at $(x_2,\,y_2,\,z_2)$,
the probability for detectors to register something would be significant
only when $(0,\,0,\,0)$, $(x_1,\,y_1,\,z_1)$ and $(x_2,\,y_2,\,z_2)$ are on a straight line,
and this is why the droplets form a straight line in a cloud chamber.

Therefore, to fully answer this question, there will be a need to specify exactly how we measure
the state after the scattering.
This is also related to what we discussed in \text{A\ref{Q:sick}}, where we mentioned that
 the transformed state looks completely normal
 as long as we measure the current $J$ and the energy-momentum tensor $T$,
 and the strange issue arises only when we try to compute non-conserved quantities
 such as the total number of original (anti-)fermions.}

\noindent The authors would like to come back to some of these issues in a not-so-distant future.

\section*{Acknowledgements}
The authors thank Simeon Hellerman and Kantaro Ohmori for valuable discussions. 
YT is supported in part by JSPS KAKENHI Grant-in-Aid (Kiban-C), No.24K06883.
KT is supported in part by the Forefront Physics and Mathematics Program
to Drive Transformation (FoPM), a World-leading Innovative Graduate Study (WINGS)
Program at the University of Tokyo.
YT and KT are supported in part by WPI Initiative, MEXT, Japan at Kavli IPMU, the University of Tokyo.
MW is supported by a Grant-in-Aid for JSPS Fellows No.~22KJ1777 and by a Grant-in-Aid for Early-Career Scientists No.~25K17387.

\appendix

\section{System on a circle with a single wall}
\label{app:onewall}
In the main text, we consider the system on a circle with two Maldacena-Ludwig walls,
for which the interpretation of the states is straightforward. 
In this appendix, we discuss what happens if we only insert a single Maldacena-Ludwig wall on the spatial circle.

In the single-wall case, the $so(8)$ currents satisfy a twisted boundary condition $J(x+2\pi R)= gJ(x)g^{-1}$.   
The boundary condition for the fermions $\psi(x)$ is trickier, 
in that we have $\psi(x+2\pi R)=\tilde \psi(x)$
and $\tilde\psi(x+2\pi R)=\psi(x)$, while $\psi(x)$ and $\tilde\psi(x)$ 
at the same point are non-local to each other.
It is unclear how to write down the Hilbert space in this description.

The Hilbert space can be determined in the following way. The modes of the $so(8)$ currents twisted by $g$ form
what is known as the twisted affine Lie algebra $D_4^{(2)}$ at level 1. 
Its representation theory is known,
and there are only two irreducible representations. When we use the form \eqref{g} of $g$ in acting on $8_C$,
these two irreducible representations are simply realized by the boundary conditions
\begin{equation}
\begin{array}{r@{\,}c@{\,}l}
\psi_C^{i=1,\ldots,7}(x+2\pi R) &=& - \psi_C^{i=1,\ldots,7}(x),\\
\psi_C^8(x+2\pi R)&=&+\psi_C^8(x),
\end{array}
\quad\text{or}\quad
\begin{array}{r@{\,}c@{\,}l}
\psi_C^{i=1,\ldots,7}(x+2\pi R) &=& + \psi_C^{i=1,\ldots,7}(x),\\
\psi_C^8(x+2\pi R)&=&-\psi_C^8(x).
\end{array}
\end{equation}
Calling the Hilbert spaces of $\psi_C(x)$ with these two boundary conditions as $\cH_{\psi_C,1}$ and $\cH_{\psi_C,2}$,
the total Hilbert space of the system with one Maldacena-Ludwig wall is given by 
\begin{equation}
\cH_{\text{one wall}}=\cH_{\psi_C,1}\oplus \cH_{\psi_C,2}.
\label{H-one-wall}
\end{equation}
Therefore, we end up having a Fock-space construction of all states, 
although $\psi_C(x)$ needs to be used for this purpose.

The states in $\cH_{\psi_C,1}$ and in $\cH_{\psi_C,2}$ transform in the vector-like and spinor-like representations of $so(7)_C$ preserved by $g$, respectively.
As both $8_V$ and $8_S$ reduce to the spinor representation of $so(7)_C$ while $8_C$ decomposes 
to the identity and the vector representation of $so(7)_C$,
we find that $\psi(x)=\psi_S(x)$ and $\tilde\psi(x)=\psi_C(x)$ maps $\cH_{\psi_C,1}$ to $\cH_{\psi_C,2}$ 
and vice versa.
In this description, it is not even clear how to explicitly write down a localized wavepacket of $\psi(x)$,
since the Fock space is made out of $\psi_C$.
This was the reason why we did not use this description in the main part of the paper.

Still, as long as the circumference of the circle is large enough compared to the size of the wavepacket,
there should not be any difference in the local behavior of the scattering by a Maldacena-Ludwig wall,
independent of whether we consider a single wall or two well-separated walls on a circle.
In principle, we only have to take the limit of an infinitely large circle, 
focusing our attention to the vicinity of a single wall, 
while letting the second wall to recede infinitely far away in the case of two walls on the circle.
However, it is not immediately clear how this independence is achieved, 
since the Hilbert space with two walls \eqref{H-two-walls} and the Hilbert space with one wall \eqref{H-one-wall} 
superficially look very different. 
This illustrates the complication in ascribing a single Hilbert space 
to continuum quantum field theories on an infinitely large space.
See also the discussion in the first comment of page~\pageref{nonsensical-philosophy} 
and the footnote \pageref{nonsensical-philosophy} on the same page.

\section{On operators of the form $\exp(\int dx\, f^a(x) J^a(x))$}
\label{app:B}
In the main part of this paper, we utilized the operators of the general form \begin{equation}
\exp\left(2\pi i\int dx f^a(x) J^a(x)\right)
\label{B1}
\end{equation} where $J^a(x)$ are the current operators and $f^a(x)$ are coefficient functions,
with $a$ being the adjoint index.
We acted these operators on  the vacuum state $\ket0$ to generate the states we wanted to analyze.

In the main part of the paper, the spatial direction was a circle, and $f^a(x)$ was a periodic real-valued function.
Then there was no question of the existence of this operator,
e.g.~because we can consider a one-parameter family \begin{equation}
\exp\left(2\pi i s\int dx f^a(x) J^a(x)\right), \qquad s\in [0,1].
\end{equation}

When $f^a\equiv f^a(x)$ is a constant independent of $x$, these operators 
are simply the global symmetry transformation $g:=\exp( 2\pi i f^a T^a) \in G$,
where $G$ is the symmetry group.
Then the operator \eqref{B1} can be thought of as a position-dependent symmetry operator $U[g(x)]$
specified by a function $g(x)=\exp(2\pi i f^a(x)T^a)\in G$.
When $f^a(x)$ is a real-valued periodic function on $S^1$, 
the function $g(x)$ defines a homotopically trivial loop in $G$,
and we just argued that the operator \eqref{B1} always exists in this case.

It is tempting, then, to think that these operators would `exist' even when the loop $g(x)$ is homotopically nontrivial on $S^1$.
Similarly, on the infinite real spatial line $\mathbb{R}$, we might be tempted to consider these operators 
when $g(x)$ tends to two different constant values when $x\to -\infty$ and $x\to +\infty$.
But the existence of these operators
and the properties of the operators when they exist are somewhat subtle questions.

For example, when $G=U(1)$ and is of level one, the loop $g(x) \in U(1)$ of winding number one corresponds
to the operator \begin{equation}
U=\exp\left(2\pi i \int_{S^1} dx\, f(x) J(x)\right)
\end{equation} where $f(x+2\pi R)=f(x)+1$. Using $J=\partial \phi$ and 
taking $f(x)$ to be a step function at $x=x_0$, this becomes the operator $e^{2\pi i\phi(x_0)}$,
which is the standard bosonization formula of a fermion $\psi(x)$.
Then, $U$ acting on the vacuum $\ket 0$ should create a state of charge one, 
which is outside of the vacuum sector.
Furthermore, such a state became automatically fermionic.
These issues when $G=U(1)$ were explored more fully e.g.~in \cite{Okada:2025kie}.

In the case of eight Majorana-Weyl fermions,  which was the main topic of this paper,
we considered operators \eqref{before} and \eqref{after} in the main text, which were 
\begin{align}
U[g_1(x)]&:=\exp\left( 2\pi i\int dx\,  c(x) J^1(x) \right) ,\\
U[g_2(x)]&:=\exp\left(2\pi i \int dx\,  c(x) \frac12 (J^1(x)+J^2(x)+J^3(x)+J^4(x)) \right) 
\end{align}
where $J^{1,2,3,4}(x)$ are current operators in the Cartan,
and $c(x)$ was one inside $[x_1,x_2]$ and was zero outside of it. 
These operators created an excitation around $x\sim x_1$
and an anti-excitation around $x\sim x_2$.

Instead, we can try to consider operators above, $U[g_{1,2}(x)]$, where $c(x)$ starts from zero 
and increases by one around $x\sim x_1$, without going back down to zero.
In this case, $g_1(x)$ is a closed loop in $SO(8)_V$ corresponding to the generator $\pi_1(SO(8)_V)=\bZ_2$.
In contrast, $g_2(x)$ is not closed and connects the identity $1\in SO(8)_V$ 
and the central element $-1\in SO(8)_V$.
Therefore, $U[g_2(x)]$ acts by $+1\in SO(8)$ to the left of $x_1$, 
and by $-1\in SO(8)$ to the right of $x_1$.
This corresponds to the fact that $\tilde\psi(x)$ is an operator in the R-sector, living at the end of the twist line.
Because of this, $U[g_1(x)]$ acts within $\cH_{\psi,\text{NS}}$ and $\cH_{\psi.\text{R}}$, respectively,
while $U[g_2(x)]$ acts by exchanging them.

Another point of view is to start from the vacuum representation $\chi_0$ of $so(8)_1$.
The symmetry group acting on $\chi_0$ is $SO(8)/\bZ_2$, and then neither $g_1(x)$ nor $g_2(x)$
gives a closed loop in it. 
Therefore, $g_{1,2}(x)$ both correspond to nontrivial loops of $\pi_1(SO(8)/\bZ_2)=\bZ_2\times \bZ_2$,
and they do not exist as an operator acting within $\chi_0$,
and exist only as operators on  a suitably extended Hilbert space, containing also $\chi_V$, $\chi_S$ or $\chi_C$.

\section{Estimate of a certain integral}
\label{app:C}
Here, we consider the case $\alpha=1/2$ and estimate the integral \begin{equation}
I :=\int_0^{2\pi}\frac{d\phi_1}{2\pi}\int_0^{2\pi}\frac{d\phi_2}{2\pi}|D_{<,<}(e^{i\phi_1},e^{i\phi_2})|^2,
\end{equation}
where $D_{<,<}(z_1,z_2)$ was given in \eqref{Ddef}.
Our aim is to show that it is bounded from above by $C(\log\frac{1}{\epsilon})^2$,
where $C$ is an $\epsilon$-independent constant.

Fix $\delta>0$ such that
\begin{equation}
    \delta<\min\left\{\frac{|\theta_2-\theta_1|}{2},\frac{\pi}{4}\right\},
\end{equation}
and take $\epsilon$ sufficiently small so that $1-e^{-\epsilon}<\delta$.
We separate the region of integration of $\phi_{1,2}$ into
\begin{equation}
    I_1=(\theta_1-\delta,\theta_1+\delta),\ I_2=(\theta_2-\delta,\theta_2+\delta),\ J=[0,2\pi)\setminus(I_1\cup I_2).
\end{equation}
For $\phi\in J$, we have
\begin{equation}
    \frac{1}{2}\delta\le|e^{i\phi}-e^{-\epsilon+i\theta_{1,2}}|\le2.
\end{equation}

\paragraph{$J\times J$:}
For $\phi_{1,2}\in J$, it can easily be seen that $|D_{<,<}(e^{i\phi_1},e^{i\phi_2})|$ is bounded above by a constant independent of $\epsilon$.

\paragraph{$I_1\times J$:}
For $\phi_1\in I_1,\phi_2\in J$, we can bound the part depending on $\phi_2$ above by a constant independent of $\epsilon$. The divergence arises from
\begin{equation}
    \int_{\theta_1-\delta}^{\theta_1+\delta}d\phi_1\,\frac{1}{|e^{i\phi_1}-e^{-\epsilon+i\theta_1}|}.
\end{equation}
Rotating $\phi_1$, it is sufficient to see the divergence of
\begin{equation}
    \int_{-\delta}^\delta d\phi\frac{1}{|e^{i\phi}-e^{-\epsilon}|}.
\end{equation}
Since $|\phi|<\delta\le\frac{\pi}{4}$, we have $1-\cos\phi\ge\frac{1}{4}\phi^2$ and
\begin{equation}
    |e^{i\phi}-e^{-\epsilon}|^2
    =(1-e^{-\epsilon})^2+2e^{-\epsilon}(1-\cos\phi)
    \ge(1-e^{-\epsilon})^2+\frac{1}{2}e^{-\epsilon}\phi^2.
\end{equation}
Then
\begin{equation}
    \int_{-\delta}^\delta d\phi\,\frac{1}{|e^{i\phi}-e^{-\epsilon}|}
    \le\int_{-\delta}^\delta d\phi\,\frac{1}{\sqrt{(1-e^{-\epsilon})^2+\frac{1}{2}e^{-\epsilon}\phi^2}}
    =\sqrt{2e^{\epsilon}}\left[\sinh^{-1}\frac{\sqrt{e^{-\epsilon}/2}}{1-e^{-\epsilon}}\phi\right]_{-\delta}^\delta.
\end{equation}
$\sqrt{2e^{\epsilon}},\sqrt{e^{-\epsilon}/2}$ are bounded above by a constant independent of $\epsilon$. Since $\sinh^{-1}x$ is bounded from above by $\log x$ multiplied by a constant when $x$ is sufficiently large, we have
\begin{equation}
    \int_{-\delta}^\delta d\phi\,\frac{1}{|e^{i\phi}-e^{-\epsilon}|}\le C\log\frac{\delta}{1-e^{-\epsilon}}\le C\log\frac{2\delta}{\epsilon},
\end{equation}
where $C>0$ is a constant independent of $\epsilon$ and we used $1-e^{-\epsilon}\ge\frac{\epsilon}{2}$ for $0<\epsilon<1$.

A similar argument applies to the cases $I_2\times J,J\times I_1,J\times I_2$.

\paragraph{$I_1\times I_1$:}
For $\phi_{1,2}\in I_1$, we can approximate $D_{<,<}(z_1,z_2)$ by $D_{<,<}(z,z)$ and in
\begin{equation}
    |D_{<,<}(z,z)|\le\frac{1}{2}\left(\frac{1}{|z-e^{-\epsilon+i\theta_1}|}+\frac{1}{|z-e^{-\epsilon+i\theta_2}|}\right),
\end{equation}
the $\frac{1}{|z-e^{-\epsilon+i\theta_1}|}$ term diverges. Therefore, the divergence is roughly given by
\begin{equation}
    \int_{\theta_1-\delta}^{\theta_1+\delta}d\phi_1\,\frac{1}{|e^{i\phi_1}-e^{-\epsilon+i\theta_1}|}\int_{\theta_1-\delta}^{\theta_1+\delta}d\phi_2\,\frac{1}{|e^{i\phi_2}-e^{-\epsilon+i\theta_1}|},
\end{equation}
and this is $O((\log\frac{1}{\epsilon})^2)$ for the same reason as in $I_1\times J$.

A similar argument applies to the case $I_2\times I_2$.

\paragraph{$I_1\times I_2$:}
For $\phi_1\in I_1,\phi_2\in I_2$, the divergence is roughly given by
\begin{equation}
    \int_{\theta_1-\delta}^{\theta_1+\delta}d\phi_1\,\frac{1}{|e^{i\phi_1}-e^{-\epsilon+i\theta_1}|}\int_{\theta_2-\delta}^{\theta_2+\delta}d\phi_2\,\frac{1}{|e^{i\phi_2}-e^{-\epsilon+i\theta_2}|},
\end{equation}
and this is also $O((\log\frac{1}{\epsilon})^2)$. A similar argument applies to the case $I_2\times I_1$.

When $1/2<\alpha<1$, the denominator becomes $|...|^{2\alpha}$ and the divergence is no longer logarithmic.

\paragraph{Conclusion:}
From the above results, we have
\begin{equation}
    I \lesssim O\left(\left(\log\frac{1}{\epsilon}\right)^2\right).
\end{equation}

\bibliographystyle{JHEP}
\bibliography{ref}

\end{document}